%% file: main.tex
\documentclass[conference]{IEEEtran}
\IEEEoverridecommandlockouts
\pdfoutput=1 
\usepackage{cite}
\usepackage{amsmath,amssymb,amsfonts}
\usepackage{algorithmic}
\usepackage{graphicx}
\usepackage{textcomp}
\usepackage{xcolor}

\usepackage{braket}
\usepackage{bbold}

\newcommand{\unity}{1\!\!1}
\renewcommand{\unity}{\mathbb{1}}

\def\BibTeX{{\rm B\kern-.05em{\sc i\kern-.025em b}\kern-.08em
    T\kern-.1667em\lower.7ex\hbox{E}\kern-.125emX}}
\begin{document}

\title{Adiabatic Quantum Computing for Solving the Weapon Target
Assignment Problem
}

\author{\IEEEauthorblockN{Veit Stoo\ss, Martin Ulmke, Felix Govaers}
\IEEEauthorblockA{Sensor Data and Information Fusion\\
Fraunhofer FKIE\\  53343 Wachtberg, Germany\\
Email: veit.stooss@fkie.fraunhofer.de, martin.ulmke@fkie.fraunhofer.de,
felix.govaers@fkie.fraunhofer.de}
}

\maketitle

\begin{abstract}
Quantum computing promises significant improvements of computation capabilities in various fields such as machine learning and complex optimization problems. Recent technological advancements suggest that the adiabatic quantum computing ansatz may soon see practical applications. In this work, we adopt this computation paradigm to develop a quantum computation based solver of the well--known weapon target assignment problem, an NP-hard  nonlinear integer programming optimization task. The feasibility of the presented model is demonstrated by numerical simulation of the adiabatic evolution of a system of quantum bits towards the optimal solution encoded in the model Hamiltonian. Over all, the described method is not limited to the context of weapon management but is, with slight modifications to the model Hamiltonian, applicable to worker-task allocation optimization in general.
\end{abstract}


\input{introduction}
\input{problem}
\input{quantum-formulation}
\input{ising-formulation}

\input{numerics}
\input{summary}

\bibliographystyle{IEEEtran}
\bibliography{references}

\end{document}

%% file: introduction.tex
\section{Introduction}
Quantum computing is considered the next key technology which will drastically increase computation speed and scalability in a wide range of optimization tasks including applications in machine learning \cite{Schuld2014}. 
Besides the quantum gate programming paradigm for general purpose quantum computers and its physical implementations \cite{Jazaeri2019,huang2020superconducting}, sig\-nificant progress is made in the field of adiabatic quantum computing \cite{RevModPhys.90.015002, Bian2010, Johnson2011} making commercial devices available for applications to various optimization problems. However, due to the underlying physical structure of adiabatic quantum computation hardware, they are best suited to solve quadratic unconstrained binary optimization problems or Ising models. It is thus necessary to reformulate and/or approximate given optimization objectives in terms of an Ising model, as it is done for example in \cite{Bauckhage2017} for binary clustering.\\

The so--called weapon target assignment (WTA) problem is an optimization task that draws great interest in the field of operations research \cite{KLINE2019226}. It is an optimization task, which deals with the general issue of optimized assignment of $m$ weapons or workers to $n$ targets or tasks, based on the probabilities of successful task completion and the (threat) value of the given targets or tasks. It can be characterized as a non-linear integer programming problem and belongs to the class of NP-complete computational complexity.\\

The goal of this paper is to derive a model of the weapon target assignment problem that can be executed on state of the art adiabatic quantum processing hardware. 
The paper is therefore structured as follows: In Section \ref{Sec:problem}, the problem formulation of target based weapon target assignment is introduced in more detail. In Section \ref{Sec:quantum}, we present the derivation of an equivalent formulation using a spin basis and Pauli operators on this basis. In Section \ref{Sec:ising}, an Ising-type Hamiltonian only containing two-spin interaction terms and self energies is constructed to reflect the target function attributes of the WTA problem. Finally, in Section \ref{Sec:numerics}, numeric simulations of a minimal practical example are presented.

%% file: problem.tex
\section{Problem Description and Notation}
\label{Sec:problem}
The WTA problem draws great interest in the field of operations or command and control research in the context of the so-called OODA (observe, orient, decide, act) loop of command systems. There exist several formulations of the optimization objective which are based either on survivability of defended assets (asset based), destruction of attacking targets (target based) \cite{Zhang2020}. Furthermore a single time step (static WTA) or a dynamically changing scenario (dynamic WTA) may be considered. Owed to the nature of the optimization method which basically searches for minimum energies of a physical system, this work discusses a target based formulation where the goal is to minimize the value of surviving attackers and thus the threat value they represent. Furthermore, for simplicity the static weapon target assignment (SWTA) problem is considered.
The target function of this problem formulation is given by:
\begin{align}
T=\sum_{j=1}^n V_j \prod_{i=1}^m (1-p_{ij})^{x_{ij}}
\label{eq:wta}
\end{align}
with $i=1, \dots, m$ weapons, $j=1, \dots, n$ targets, $V_j$ the target threat value (penalty for non-completion of task), and $p_{ij} \in [0,1]$ the destruction probability if weapon $i$ engages target $j$. The $N=mn$ entries of the $x_{ij}$ integer valued decision matrix indicate which engagements are taken (1: chosen engagement, 0: not chosen engagement).
The goal is now to choose from the solution space $\mathcal{G}$ with $|\mathcal{G}|=2^{N}$ possible solutions the member which minimizes $T$ \eqref{eq:wta}:
\begin{align}
S= \min_{x_{ij}\in \mathcal{G}} T (x_{ij}) 
\label{eq:wtamin}
\end{align}
under the constraints
\begin{align}
\sum_{j=1}^n x_{ij} \le 1 \quad \forall \, i
\label{eq:wtabc}
\end{align}
assuming that each weapon can be assigned to at most one target.

In the current literature \cite{Hosein1990,Ahuja2007,Ma2015,Li2017,Sonuc2017,Zhang2020} several classical algorithms are developed to solve this assignment problem in an (approximately) optimal way. Among the methods are sequential heuristic assignments based on marginal return values and stochastic methods like genetic algorithms, ant-colony algorithms and simulated annealing. 
To our knowledge however, there are no methods which leverage the emerging capabilities of quantum computing paradigms. Especially quantum annealing can yield improved scalability to large scenarios and significant speedups in finding global optima of optimization problems if the solution to a given problem is properly encoded in the ground state of a Hamilton operator.
We therefore aim to reformulate the WTA problem stated in \eqref{eq:wta} for computation on current spin based adiabatic quantum computers, which is presented in the following chapter.

%% file: quantum-formulation.tex
\section{Quantum formulation of the WTA problem based on interacting spins}
\label{Sec:quantum}



The solution which needs to be calculated in WTA optimization is the decision matrix $x_{ij}$. The idea of the adiabatic quantum computing ansatz to WTA (AQC WTA) is to represent each entry of $x_{ij}$ with the state of a physical 2-level spin system, denoted as $\ket{s_{ij}}$ in Dirac notation, as they are used as qubits in current quantum hardware. $\ket{s_{ij}}$ is a spin$-\frac{1}{2}$ quantum state with basis states 
\begin{align}
 \ket{0} = \begin{pmatrix} 1 \\ 0 \end{pmatrix} \quad \text{and} \quad
 \ket{1} = \begin{pmatrix} 0 \\ 1 \end{pmatrix} 
\end{align}
which are eigenstates of the Pauli-operator 
\begin{align}
\sigma_z  =
\begin{pmatrix}
1 & \phantom{-}0 \\
0 & -1 
\end{pmatrix}
\end{align}

with eigenvalues 1 and -1, respectively.
Thus, they are also eigenstates of the operator $\sigma_s  = \frac{1}{2} ( \unity - \sigma_z)$ with 
\begin{align}
\sigma_s  \ket{s} = s \ket{s} 
\end{align}
for $s \in \{ 0, 1\}$.

The full quantum state can be expressed as a product state of $N=m n$ coupled qubits over the Hilbert spaces of each individual qubit with the correspondence given above: 
\begin{align}
\ket{S} = \ket{s_{11}} \otimes \dots \otimes\ket{s_{mn}}  \equiv \ket{ s_{11}, \dots , s_{mn}} .
\end{align}


There exist $K=2^{N}$ basis states of the systems of $N$ spins, denoted as $\ket{S_k}$ ($k=1, \dots , K$),
and a general time dependent state of the coupled qubit system can be written as a superposition of the basis states with dynamic state coefficients $c_k (t)$:  
\begin{align}
\ket{\psi(t)} = \sum_{k=1}^K c_k (t)  \ket{S_k} .
\label{eq:state}
\end{align}

Under the influence of a time-dependent Hamiltonian $H(t)$ the behavior of the system described in \eqref{eq:state} is determined by the time-dependent Schr\"odinger equation

\begin{align}
\frac{\partial}{\partial t}\ket{\psi(t)} = -\imath\hbar H(t)\ket{\psi(t)} .
\label{eq:tdse}
\end{align}

In adiabatic quantum computation the time dependent Hamiltonian is constructed such that it moves from a problem independent Hamilton operator with known ground state to the Hamilton operator which encodes the solution of the optimization problem as its ground state. The adiabatic theorem \cite{Born1928} states that if the system starts in the ground state and the time dependent change only happens gradually, then the system will remain in the ground state of the final Hamilton operator \cite{Farhi2000,Lucas2014,RevModPhys.90.015002}. A state measurement can thus yield the optimized solution.
Following the guidelines in \cite{Bian2010,Farhi2000,Lucas2014}, the target function \eqref{eq:wta} of the optimization problem is treated similar to a satisfiability problem, where the conditions are not boolean but numeric values, which need to be minimized simultaneously. 
Due to the structure of $T$, each term of the sum can be treated as its own condition, which has to be optimized. 
The Hamiltonian is derived by defining the spin operator $\sigma^{ij}_s$  on the subspace of each qubit as
\begin{align}
\hat \sigma_s^{ij} = \unity^{1,1}  \otimes \unity^{1,2}  \otimes \dots \otimes \sigma_s^{ij} \otimes\ \dots \otimes \unity^{nm} . 
\end{align}

In the following, we will not discriminate between $\hat \sigma_s^{ij}$ and $\sigma_s^{ij}$ if the context is clear.

We can use this to rewrite $T$ to get the target state Hamiltonian
\begin{align}
H_F= \sum_{j=1}^n V_j \prod_{i=1}^m (1-p_{ij} \, \sigma_s^{ij} ) .
\label{eq:finalh}
\end{align}

Here, the product terms behave exactly like the classical formulation with the integer decision matrix entries, i.e.\  the terms are either 1 for $x_{ij}=0$ and $\ket{s_{ij}} = \ket 0$ or $(1-p_{ij})$ for $x_{ij}=1$ and $\ket{s_{ij}} = \ket 1$. 
In addition, boundary conditions such as limitations on weapon assignments \eqref{eq:wtabc} can be imposed using large energy penalties if the condition is not fulfilled.

However, for more than two weapons (i.e.\ $m>2$) the Hamiltonian as stated above can no longer be used on current quantum hardware as it contains higher order non-linear spin coupling terms (powers of $\sigma_s$ greater than 2). Current hardware can only process problems  that can be cast into the form of spin Hamiltonians with quadratic interactions, e.g., the widely used Ising model, which has the form
\begin{align}
H_I = - \sum_{i,j} J_{ij} \sigma_z^i \sigma_z^j - \mu \sum_j h_j \sigma_z^j ,
\end{align}

where $\mu$ represents the magnetic moment of the spin, $J_{ij}$ the interaction strength between all pairs of sites and $h_j$ a local magnetic field on spin at site j. 
The parameters $J_{ij}$ and $h_j$ of the model can be set on the hardware side, which enables inputting an optimization problem into a quantum processor if the problem can be formulated in terms of such an Ising-type Hamiltonian.

%% file: ising-formulation.tex
\section{Ising-type Reformulation of the WTA Hamiltonian with Spin-Spin Interactions}
\label{Sec:ising}

In order to calculate solutions to the WTA problem on adiabatic quantum computing hardware based on spin interactions, $H_F$ \eqref{eq:finalh} has to be cast into the form of an Ising-type Hamiltonian. Because of the higher-order couplings this cannot be done in an exact way. Therefore, we construct a second order approximation including boundary conditions in a heuristic way.

The construction argument is that spins only need to influence spins in the same row $(j=1, \dots , n)$ and the same column $(i=1, \dots , m)$. This is based on the structure of the WTA problem and target function $T$ \eqref{eq:wta}. Each weapon can be used only on one target therefore each row $i$ of $x_{ij}$ needs to be optimized separately and simultaneously, while each column represents the decision how many weapons should be used on one target, depending on target value $V_j$ and the overall destruction probabilities. Thus, decision entries  that are not in the same column or the same row cannot influence each other directly. We, therefore, only have to consider spin interaction terms within the same row or the same column. The idea is now to set up a WTA Hamiltonian, which is only of second order in $\sigma_z$ but reflects the behavior of the full classical problem formulation.

The building blocks for the overall Hamiltonian are interactions between two spins within a row and within a column. 
%
%

The first block represents the case of a single weapon $i$ and two targets $j,j'$, which gives inserted in $H_F$ (see \eqref{eq:finalh}):
\begin{align}
H_\text{row}(i;j,j') =  V_j (1-p_{ij} \, \sigma_s^{ij} ) +  V_{j'} (1-p_{ij'} \, \sigma_s^{ij'} ) .
\end{align} 

The boundary conditions \eqref{eq:wtabc} can be imposed by the Hamiltonian:
\begin{align}
H_\text{bc}(i;j,j') & =  C \, \sigma_s^{ij} \sigma_s^{ij'}  \\ 
& =\frac{C}{4} \left(1 -\sigma_z^{ij} -\sigma_z^{ij'}+\sigma_z^{ij}\sigma_z^{ij'}\right) .
\end{align} 

$H_\text{bc}$ punishes each pair of double assignments with an energy term $C$.

The second block, representing the case of a single target and two weapons, gives inserted in $H_F$:  
\begin{align}
H_\text{col}(j;i,i') = V_j  (1-p_{ij} \, \sigma_s^{ij} ) (1-p_{i'j} \, \sigma_s^{i'j} ) .
\end{align} 

Using these blocks, a problem Hamiltonian of arbitrary size can be constructed as long as all pairwise interactions within each row and each column are included and summed over. 
\begin{align}
\tilde H_F & = \sum_{i=1}^m \sum_{j'<j=1}^n  H_\text{row}(i;j,j') + H_\text{bc}(i;j,j') \nonumber \\ 
& + \sum_{j=1}^n \sum_{i'<i=1}^m H_\text{col}(j;i,i') .
\label{eq:hfising}
\end{align} 

In contrast to the original Hamiltonian \eqref{eq:finalh}, $\tilde H_F$ is of second order in $\sigma_s^{ij}$ and, hence, in $\sigma_z^{ij}$. 
The overall lowest (ground state) energy of this Hamiltonian encodes the solution of the assignment problem. 

Next, the general procedure for the application of adiabatic evolution of the system discussed in the previous section is carried out. For the initial state Hamiltonian $H_B$ and the initial ground state $\ket{\Psi(0)}$ we use the standard form that is prescribed in \cite{Farhi2000,Bauckhage2017}.
\begin{align}
H_B & = - \sum_{ij} \sigma_x^{ij} \quad \text{with} \quad  \sigma_x = \begin{pmatrix} 0 & 1 \\ 1 & 0 \end{pmatrix} \label{eq:hbising} \\
\ket{\Psi(0)} & = \sum_{s_{11}=0}^1 \dots \sum_{s_{mn}=0}^1 \ket{s_{11}, \dots , s_{mn}} .
\end{align} 

Following \cite{Lucas2014}, the adiabatic evolution is governed by the time dependent Hamiltonian:
\begin{eqnarray}
H(t) = \left( 1 - \frac{t}{T_a}\right) H_B + \frac{t}{T_a} \, \tilde H_F
\end{eqnarray} 

with the characteristic duration $T_a$ chosen suitably long such that during the adiabatic evolution between the two system states most of the wave function remains in the ground state. Using this Hamiltonian,  the time evolution of the system is calculated using the QuTiP package \cite{Johansson2013} in Python, which solves the time dependent Schrödinger equation
\begin{align}
\ket{\Psi(t)} = -i \int_0^t H(t') \ket{\Psi(t')} dt'
\end{align} 
numerically using the Master equation where the states and operators are given in matrix representations.

%% file: numerics.tex
\section{Numeric Simulations}
\label{Sec:numerics}

In order to test the validity of the proposed optimization model, we consider three systems with a 12-qubit register each. This number of qubits represents the computational limit of the hardware given by the amount of RAM needed to represent the state basis and operators. The systems treat the cases of (1) 4 weapons and 3 targets, (2) 6 weapons and 2 targets and (3) 2 weapons and 6 targets. The target threat values are chosen w.o.l.g. on the order of 1. For all examples the destruction probabilities are chosen such that at least one of the randomized values yields a very large destruction probability. The Hamiltonian operators are constructed for each system according to \eqref{eq:hfising} and \eqref{eq:hbising}.

\begin{figure}[htbp]
\centerline{\includegraphics[width=0.5\textwidth]{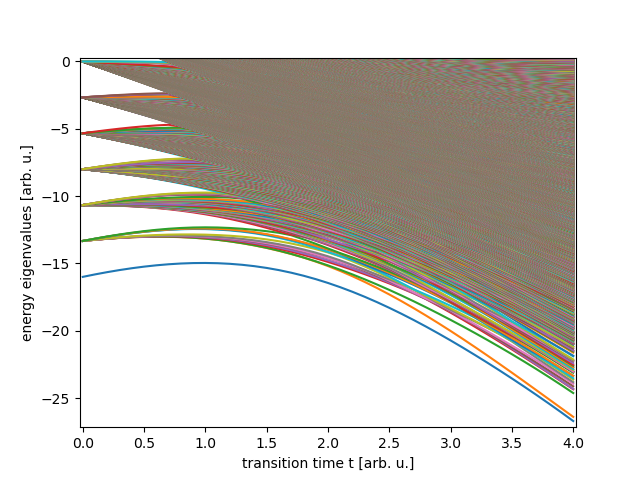}}
\caption{Energy eigenvalues over time of the 4x3 system. A finite gap between the ground state and all exicted states is present at all times.}
\label{fig:ex1a}
\end{figure}

\begin{figure}[htbp]
\centerline{\includegraphics[width=0.5\textwidth]{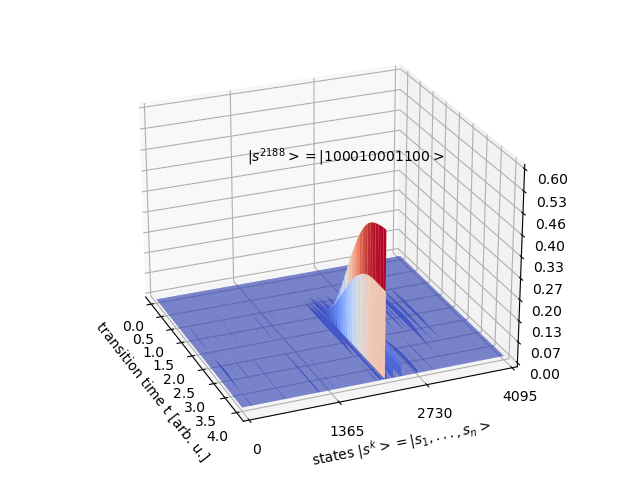}}
\caption{Time evolution of state population probabilities over the adiabatic transition process in the 4x3 system. The final state shows a distinct maximum at the position of the optimal solution}
\label{fig:ex1b}
\end{figure}

\begin{figure}[htbp]
\centerline{\includegraphics[width=0.5\textwidth]{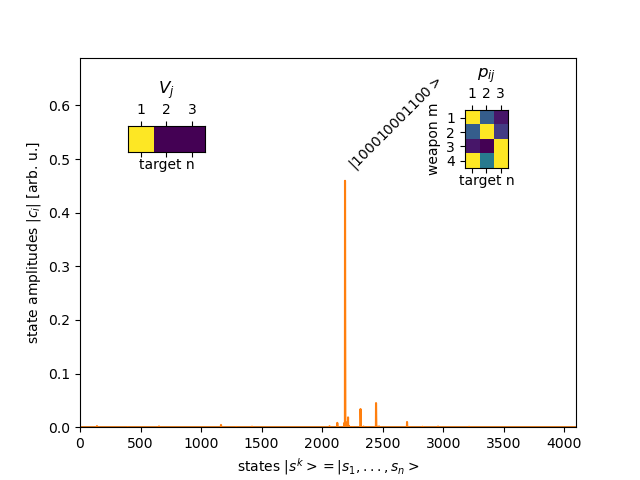}}
\caption{Detailed evaluation of the final state amplitudes. The maximum position transformed to binary representation represents the spin state of the optimal solution. Insets give the threat values and destruction probabilities of possible assignments as heat maps (blue: low values, yellow: high values). The boundary conditions are fullfilled and the algorithm selects the engagements with highest destruction probabilities.}
\label{fig:ex1c}
\end{figure}

First we present a balanced engagement in system (1). Fig.~\ref{fig:ex1a} shows the evolution of the eigenenergies of $H(t)$ during the adiabatic transition. Note, that at each time there is a finite gap between the ground state and the first and higher excited states guaranteeing that the adiabatic evolution is viable. In Fig.~\ref{fig:ex1b} the state amplitude evolution during the adiabatic transition to the final Hamiltonian $H_F$ is shown. Starting from a state with equally distributed state populations the final state shows clear shifts of probabilitiy towards favorable solution states, with the maximum at the position of the optimal solution. By taking the position of the maximum, the spin state of the quantum register can be recovered easily due to the ordering of the state basis. The basis states are ordered as ascending integers in binary representation. Thus, a given spin state translates to the position $k_{sol}$ of the basis state $\ket{S_{k}}$ according to
\begin{align}
k_{sol}=\sum_{i=1}^m \sum_{j=1}^n s_{ij} \, 2^{m(j-1)+i-1} . 
\end{align}
Consequently, the inverse transformation to binary can construct the solution from the position of the basis state. Looking at the insets of Fig.~\ref{fig:ex1c}, which show the threat values $V_j$ and destruction probabilities $p_{ij}$ (blue: low values, yellow: high values), it can be seen, that the optimization indeed picked the correct solution of the optimization. The correct convergence was furthermore validated by a separate calculation using the cross entropy (CE) statistical optimization method \cite{Rubinstein1999} for the given system, which yields the same result.

\begin{figure}[htbp]
\centerline{\includegraphics[width=0.5\textwidth]{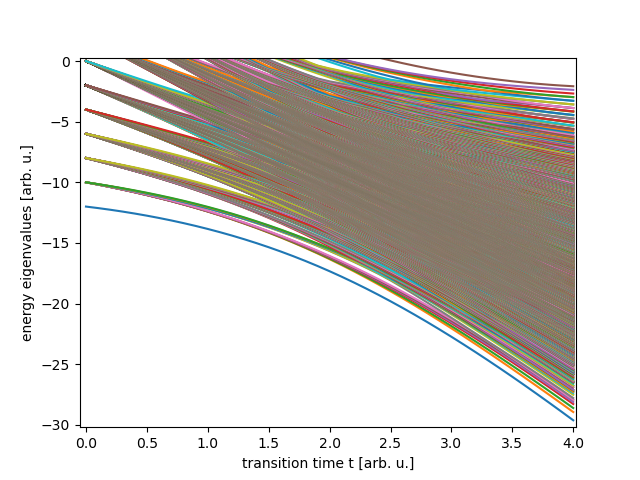}}
\caption{Energy eigenvalues over time of the 6x2 system. A finite gap between the ground state and all exicted states is present at all times.}
\label{fig:ex2a}
\end{figure}

\begin{figure}[htbp]
\centerline{\includegraphics[width=0.5\textwidth]{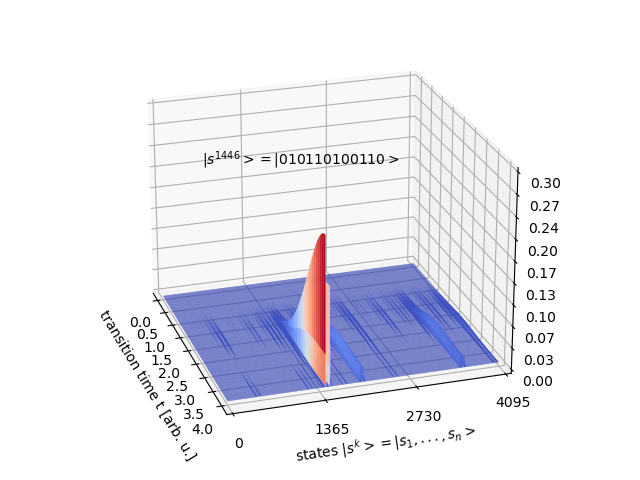}}
\caption{Time evolution of state population probabilities over the adiabatic transition process in the 6x2 system. The final state shows a distinct maximum at the position of the optimal solution.}
\label{fig:ex2b}
\end{figure}

\begin{figure}[htbp]
\centerline{\includegraphics[width=0.5\textwidth]{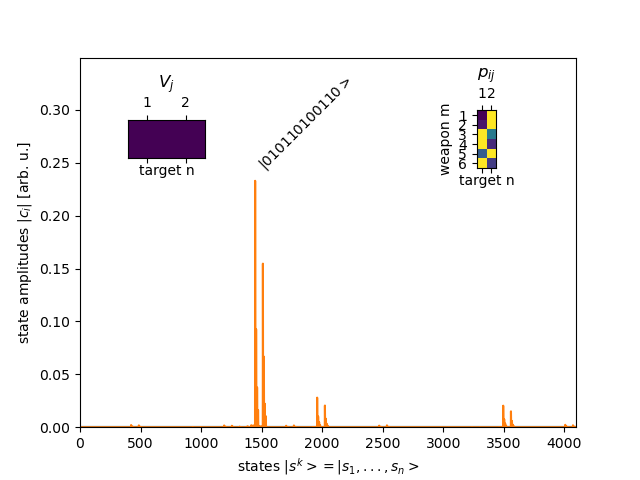}}
\caption{Same as Fig.~\ref{fig:ex1c} but for the 6x2-system.}
\label{fig:ex2c}
\end{figure}

\begin{figure}[htbp]
\centerline{\includegraphics[width=0.5\textwidth]{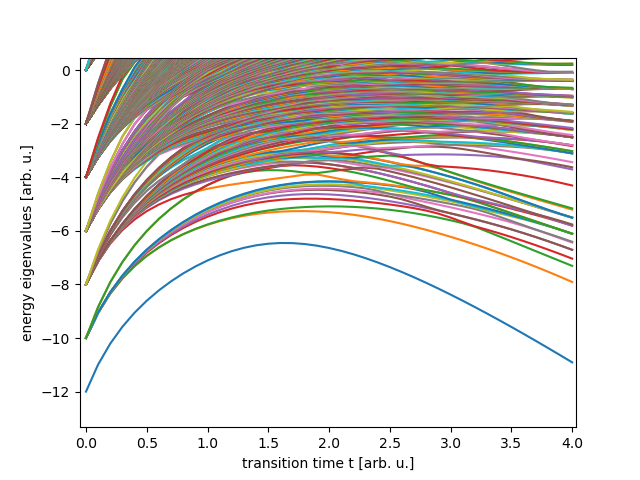}}
\caption{Energy eigenvalues over time of the 2x6 system. A finite gap between the ground state and all exicted states is present at all times.}
\label{fig:ex3a}
\end{figure}

\begin{figure}[htbp]
\centerline{\includegraphics[width=0.5\textwidth]{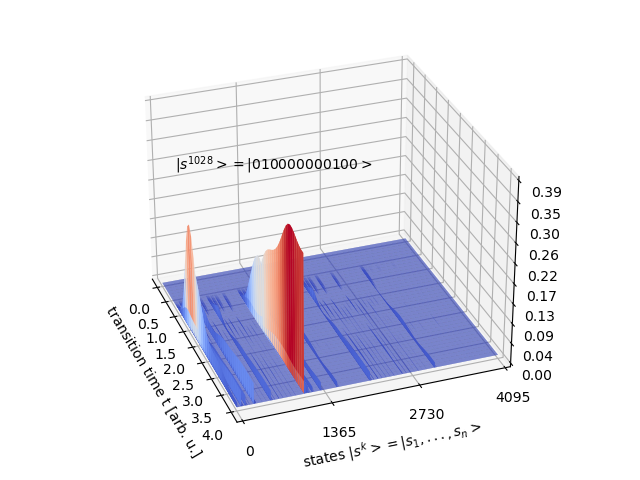}}
\caption{Time evolution of state population probabilities over the adiabatic transition process in the 2x6 system. The final state shows a distinct maximum at the position of the optimal solution}
\label{fig:ex3b}
\end{figure}

\begin{figure}[htbp]
\centerline{\includegraphics[width=0.5\textwidth]{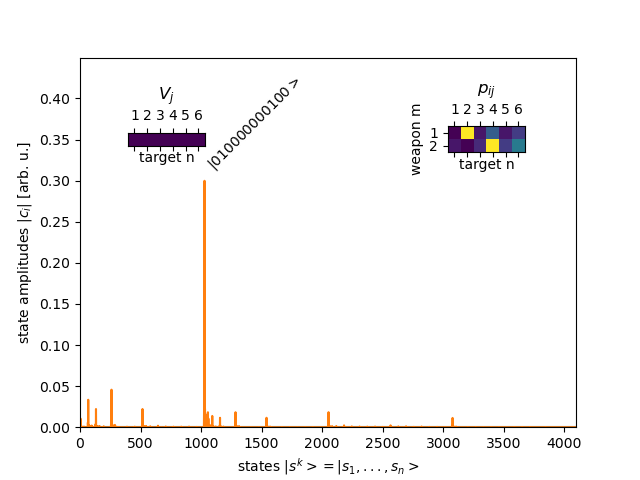}}
\caption{Same as Fig.~\ref{fig:ex1c} but for the 2x6-system.}
\label{fig:ex3c}
\end{figure}

The second system (2) represents an imbalanced engagement with superior numbers of weapons. Here, each weapon should select the target with highest destruction probability, which is indeed the case as the evaluation results in figures \ref{fig:ex2a}-\ref{fig:ex2c} show. 

The last system (3) deals with the reverse case of a superior number of targets, which tests the accurate adherence of the optimization to the imposed boundary condition, that each weapon can only engage one target simultaneously. As Figs.~\ref{fig:ex3a}-\ref{fig:ex3c} show, 
the adiabatic optimization algorithm assigns only one target per weapon and chooses the targets with highest destruction propability. The correct convergence was again checked using separate CE optimization calculations, which agree with the presented results.

%% file: summary.tex
\section{Summary and Conclusions}
\label{Sec:summary}
In summary, we showed how the target based formulation of the WTA objective can be reformulated as a second order approximation of the target function in the form of an Ising model Hamiltonian. This in turn can be implemented in adiabatic quantum computation hardware to efficiently solve the WTA problem. We showed the correct convergence of the proposed algorithm to the optimal solution for different WTA scenarios. The scenario size was only limited by the computational capabilities of the available classical hardware. In future work, it is necessary to mathematically prove that the approximate formulation indeed converges to the solution of the full WTA target function. Furthermore, with increasing capabilities of quantum computation hardware the method could be also applied to solve the dynamic weapon target assignment problem which will pose a challenge concerning the implementation of suitable boundary conditions for engagement durations but in principle only extends the solution space $x_{ij}$ by an additional time dimension. Additionally more complex formulations like asset based WTA can be considered. In any case AQC WTA promises fast solution of the assignment optimization with improved scalability to large problem instances making real time applications feasible. \\

%% file: main.bbl
\begin{thebibliography}{10}
\providecommand{\url}[1]{#1}
\csname url@samestyle\endcsname
\providecommand{\newblock}{\relax}
\providecommand{\bibinfo}[2]{#2}
\providecommand{\BIBentrySTDinterwordspacing}{\spaceskip=0pt\relax}
\providecommand{\BIBentryALTinterwordstretchfactor}{4}
\providecommand{\BIBentryALTinterwordspacing}{\spaceskip=\fontdimen2\font plus
\BIBentryALTinterwordstretchfactor\fontdimen3\font minus
  \fontdimen4\font\relax}
\providecommand{\BIBforeignlanguage}[2]{{%
\expandafter\ifx\csname l@#1\endcsname\relax
\typeout{** WARNING: IEEEtran.bst: No hyphenation pattern has been}%
\typeout{** loaded for the language `#1'. Using the pattern for}%
\typeout{** the default language instead.}%
\else
\language=\csname l@#1\endcsname
\fi
#2}}
\providecommand{\BIBdecl}{\relax}
\BIBdecl

\bibitem{Schuld2014}
I.~S. M.~Schuld and F.~Petruccione, ``An introduction to quantum machine
  learning,'' \emph{Contemporary Physics}, vol.~56, no.~2, 2014.

\bibitem{Jazaeri2019}
F.~{Jazaeri}, A.~{Beckers}, A.~{Tajalli}, and J.~{Sallese}, ``A review on
  quantum computing: From qubits to front-end electronics and cryogenic
  {MOSFET} physics,'' in \emph{2019 MIXDES - 26th International Conference
  "Mixed Design of Integrated Circuits and Systems"}, 2019, pp. 15--25.

\bibitem{huang2020superconducting}
\BIBentryALTinterwordspacing
H.-L. Huang, D.~Wu, D.~Fan, and X.~Zhu, ``Superconducting quantum computing: A
  review,'' 2020. [Online]. Available: \url{https://arxiv.org/abs/2006.10433}
\BIBentrySTDinterwordspacing

\bibitem{RevModPhys.90.015002}
\BIBentryALTinterwordspacing
T.~Albash and D.~A. Lidar, ``Adiabatic quantum computation,'' \emph{Rev. Mod.
  Phys.}, vol.~90, p. 015002, Jan 2018. [Online]. Available:
  \url{https://link.aps.org/doi/10.1103/RevModPhys.90.015002}
\BIBentrySTDinterwordspacing

\bibitem{Bian2010}
Z.~Bian, F.~A. Chudak, W.~Macready, and G.~Rose, ``The {Ising} model: teaching
  an old problem new tricks,'' D-Wave Systems, Tech. Rep., 2010.

\bibitem{Johnson2011}
M.~{Johnson et al.}, ``Quantum annealing with manufactured spins,''
  \emph{Nature}, vol. 473, no. 7346, 2011.

\bibitem{Bauckhage2017}
\BIBentryALTinterwordspacing
C.~Bauckhage, E.~Brito, K.~Cvejoski, C.~Ojeda, R.~Sifa, and S.~Wrobel,
  ``Adiabatic quantum computing for binary clustering,'' 2017. [Online].
  Available: \url{https://arxiv.org/abs/1706.05528}
\BIBentrySTDinterwordspacing

\bibitem{KLINE2019226}
A.~Kline, D.~Ahner, and R.~Hill, ``The weapon-target assignment problem,''
  \emph{Computers \& Operations Research}, vol. 105, pp. 226--236, 2019.

\bibitem{Zhang2020}
\BIBentryALTinterwordspacing
K.~Zhang, D.~Zhou, Z.~Yang, Y.~Zhao, and W.~Kong, ``Efficient decision
  approaches for asset-based dynamic weapon target assignment by a receding
  horizon and marginal return heuristic,'' \emph{Electronics}, vol.~9, no.~9,
  2020. [Online]. Available: \url{https://www.mdpi.com/2079-9292/9/9/1511}
\BIBentrySTDinterwordspacing

\bibitem{Hosein1990}
P.~A. Hosein and M.~Athans, ``Some analytical results for the dynamic
  weapon-target allocation problem,'' Defence Technical Information Center,
  Tech. Rep. ADA219281, 1990.

\bibitem{Ahuja2007}
R.~K. Ahuja, A.~Kumar, K.~C. Jha, and J.~B. Orlin, ``Exact and heuristic
  algorithms for the weapon-target assignment problem,'' \emph{Operations
  Research}, vol.~55, no.~6, pp. 1136--1146, 2007.

\bibitem{Ma2015}
F.~{Ma}, M.~{Ni}, B.~{Gao}, and Z.~{Yu}, ``An efficient algorithm for the
  weapon target assignment problem,'' in \emph{2015 IEEE International
  Conference on Information and Automation}, Aug 2015, pp. 2093--2097.

\bibitem{Li2017}
{Juan Li}, J.~{Chen}, B.~{Xin}, and {Lu Chen}, ``Efficient multi-objective
  evolutionary algorithms for solving the multi-stage weapon target assignment
  problem: A comparison study,'' in \emph{2017 IEEE Congress on Evolutionary
  Computation (CEC)}, 2017, pp. 435--442.

\bibitem{Sonuc2017}
E.~Sonuç, B.~Sen, and S.~BAYIR, ``A parallel simulated annealing algorithm for
  weapon-target assignment problem,'' \emph{International Journal of Advanced
  Computer Science and Applications}, vol.~8, 04 2017.

\bibitem{Born1928}
M.~Born and V.~Fock, ``Beweis des {A}diabatensatzes.'' \emph{Z. Physik},
  vol.~51, pp. 165--180, 08 1928.

\bibitem{Farhi2000}
\BIBentryALTinterwordspacing
E.~Farhi, J.~Goldstone, S.~Gutmann, and M.~Sipser, ``Quantum computation by
  adiabatic evolution,'' \emph{arXiv: Quantum Physics}, 2000. [Online].
  Available: \url{https://arxiv.org/abs/quant-ph/0001106}
\BIBentrySTDinterwordspacing

\bibitem{Lucas2014}
\BIBentryALTinterwordspacing
A.~Lucas, ``Ising formulations of many np problems,'' \emph{Frontiers in
  Physics}, vol.~2, 2014. [Online]. Available:
  \url{http://dx.doi.org/10.3389/fphy.2014.00005}
\BIBentrySTDinterwordspacing

\bibitem{Johansson2013}
\BIBentryALTinterwordspacing
J.~Johansson, P.~Nation, and F.~Nori, ``Qutip 2: A python framework for the
  dynamics of open quantum systems,'' \emph{Computer Physics Communications},
  vol. 184, no.~4, p. 1234–1240, Apr 2013. [Online]. Available:
  \url{http://dx.doi.org/10.1016/j.cpc.2012.11.019}
\BIBentrySTDinterwordspacing

\bibitem{Rubinstein1999}
R.~Rubinstein, ``The cross-entropy method for combinatorial and continuous
  optimization,'' \emph{Methodology and Computing in Applied Probability},
  vol.~1, p. 127–190, 1999.

\end{thebibliography}
